\documentclass[aps,pra,twocolumn,letterpaper]{revtex4}

\usepackage{amsfonts, amssymb, amsmath, graphicx, bbold}

\DeclareFontFamily{OT1}{pzc}{}
\DeclareFontShape{OT1}{pzc}{m}{it}{<-> s * [1.30] pzcmi7t}{}
\DeclareMathAlphabet{\mathpzc}{OT1}{pzc}{m}{it}
\newcommand{\vect}[1]{\boldsymbol{\mathrm{#1}}}
\newcommand{\ket}[1]{|#1\rangle}
\newcommand{\bra}[1]{\langle#1|}
\newcommand{\expval}[1]{\langle#1\rangle}
\newcommand{\abs}[1]{\left\lvert#1\right\rvert}
\newcommand{\Eqref}[1]{Eq.~\eqref{#1}}
\newcommand{\Fig}[1]{Fig.~\ref{#1}}

\def\mathi{\textrm{i}}

\def\beq{\begin{equation}}
\def\eeq{\end{equation}}
\def\bes{\begin{equation*}}
\def\ees{\end{equation*}}
\def\bfig{\begin{figure}}
\def\efig{\end{figure}}

\def\eg{e.g., }

\def\prl#1#2#3{Phys.\ Rev.\ Lett.\ {\bf #1}, #2 (#3)}
\def\pra#1#2#3{Phys.\ Rev.\ A {\bf #1}, #2 (#3)}

\def\rmp#1#2#3{Rev.\ Mod.\ Phys.\ {\bf #1}, #2 (#3)}

\begin{document}
	\title{Bose-Einstein condensates in toroidal traps: instabilities, swallow-tail loops, and self-trapping}
	\author{Soheil Baharian}
	\author{Gordon Baym}
	\affiliation{Department of Physics, University of Illinois at Urbana-Champaign, 1110 W. Green St., Urbana, IL 61801, USA}
	\date{\today}
	
	\begin{abstract}
		We study the stability and dynamics of an ultra-cold bosonic gas trapped in a toroidal geometry and driven by rotation, in the absence of dissipation. We first delineate, via the Bogoliubov mode expansion, the regions of stability and the nature of instabilities of the system for both repulsive and attractive interaction strengths. To study the response of the system to variations in the rotation rate, we introduce a ``disorder" potential, breaking the rotational symmetry. We demonstrate the breakdown of adiabaticity as the rotation rate is slowly varied and find forced tunneling between the system's eigenstates. The non-adiabaticity is signaled by the appearance of a swallow-tail loop in the lowest-energy level, a general sign of hysteresis. Then, we show that this system is in one-to-one correspondence with a trapped gas in a double-well potential and thus exhibits macroscopic quantum self-trapping. Finally, we show that self-trapping is a direct manifestation of the behavior of the lowest-energy level.
	\end{abstract}

	\maketitle
	
	\section{Introduction}
		Superfluid flow in a toroidal trap is stabilized by a large energy barrier between the current-carrying state and a state with lower angular momentum~\cite{Yang-ODLRP, AJLbook}. However, in mesoscopic systems, such as atomic Bose-Einstein condensates, the barrier can be sufficiently small that the system can tunnel quantum-mechanically to a state of lower angular momentum~\cite{MuellerGoldbartLG, CurrentStability-Kavoulakis}; furthermore, if the short-range interparticle interactions are attractive or very weakly repulsive, such a barrier does not exist, and the system can transition smoothly from the current-carrying state, to, \eg the non-rotating ground state. Recent experiments in ultra-cold bosonic systems in toroidal traps, stimulated by the possibility of shining new light on the stability and decay of supercurrents~\cite{DissipationlessFlow} as well as by possible applications in other areas, \eg interferometry~\cite{ToroidalTrap-Applications1} and atomtronics~\cite{ToroidalTrap-Applications2}, have seen such current decays~\cite{BECinToroidalTraps, BECinToroidalTraps-WeakLink, CurrentDecay-Hadzibabic}.

		The stability of superflow depends on the interparticle interactions, the rotation rate of the trap, disorder in the trapping potential, and temperature. We consider a gas of interacting bosons at zero temperature in a toroidal trap rotating at angular velocity $\Omega$ and address the question of how the single-vortex condensate with a metastable or unstable superflow evolves in the absence of dissipation, driven either by varying the rotation rate of the trap or by varying the interparticle interaction via a Feshbach resonance.

		We consider, throughout this paper, a quasi-one-dimensional gas of $N$ bosons in a thin annulus of radius $R$ and cross-sectional radius $r_0 \ll R$ at zero temperature. The basic physics of the stability can be most simply understood by considering just two single-particle levels of the annulus, the non-rotating state, $\ket{0}$, and the state with azimuthal angular momentum $\hbar$ per particle, $\ket{1}$. The Hamiltonian of this system has the familiar Nozi\`eres form~\cite{Nozieres}
		\beq
			\mathcal{H}_2 = \frac{\hbar^2}{2mR^2} N_1 + \frac{g}{2V} (N^2_0 + N^2_1 + 4 N_0 N_1)
		\eeq
		where $m$ is the particle mass, $V = 2 \pi^2 r_0^2 R$ is the volume of the annulus, $g$ is the strength of contact interactions, and $N_0$ and $N_1$ are the number of particles in $\ket{0}$ and $\ket{1}$ respectively with $N=N_0+N_1$ the total number of particles. The state with $N_1=N$ is a single-vortex state and that with $N_0=N$ is the ground state. Figure~\ref{energyLandscape}, which shows the energy per particle as a function of $N_1$ for different values of the interparticle interaction strength, illustrates the energy barrier that appears between the single-vortex state and the non-rotating ground state when $gN/V > \hbar^2/2mR^2$. With weakening interaction strength, the barrier decreases, and for $gN/V \le\hbar^2/2mR^2$, it disappears, leading to instability of the single-vortex state.
		\bfig[b]
			\centering
			\includegraphics[scale=0.165]{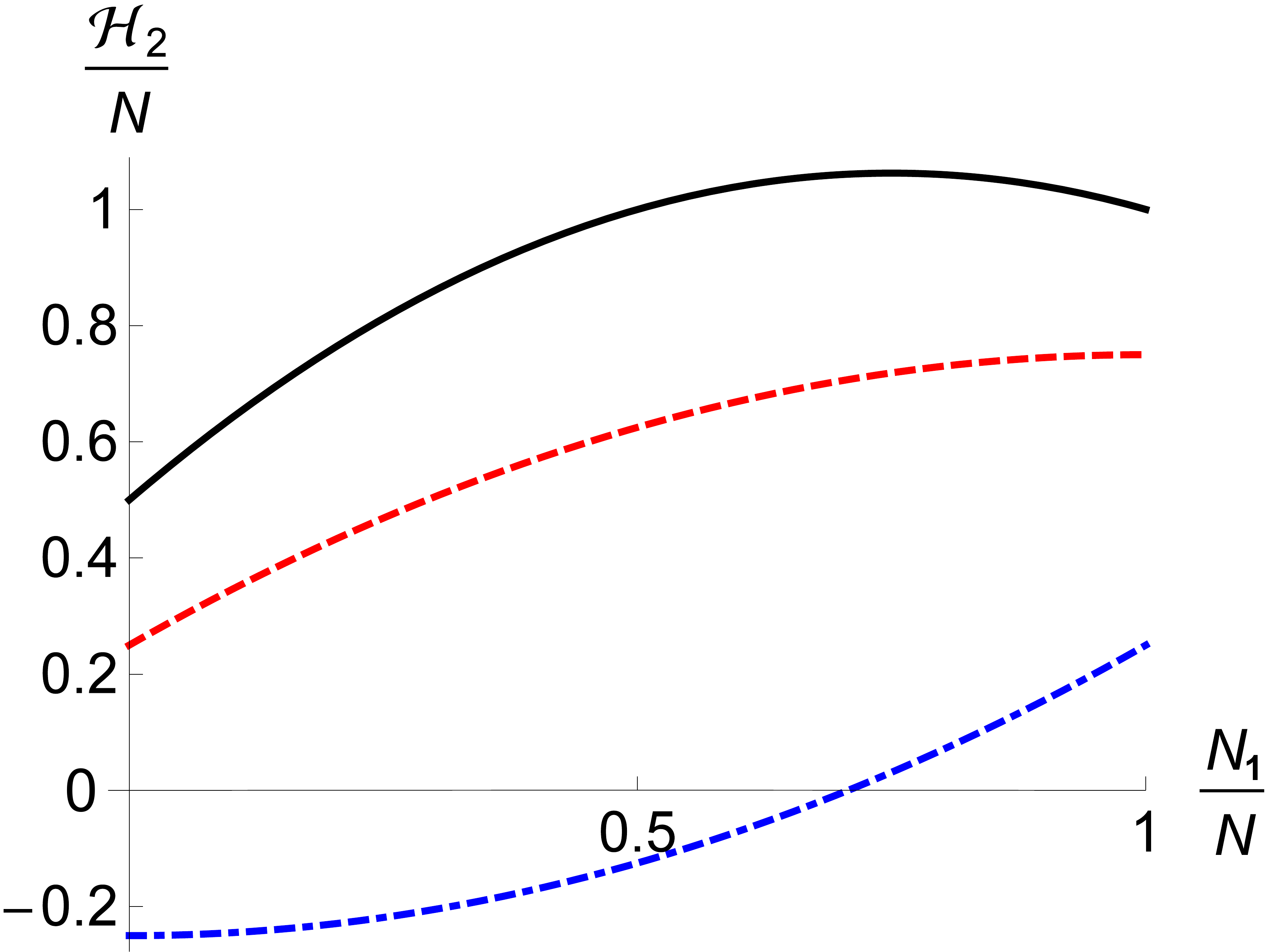}
			\caption{(Color online) Energy landscape of the two-level model as a function of the number of particles in state $\ket{1}$. Note the energy barrier between the single-vortex and ground states for $gN/V > \hbar^2/2mR^2$ (solid line, in black); for $gN/V = \hbar^2/2mR^2$ (dashed line, in red), the slope vanishes at $N_1=N$, while for $gN/V < \hbar^2/2mR^2$ (dot-dashed line, in blue) no barrier exists, indicating instability of the single-vortex state. \label{energyLandscape}}
		\efig

		We first delineate the regions of stability and the nature of instabilities of the full system as functions of the external rotation frequency of the trap, for both positive and negative interaction strengths. In general, the stability of the flow is manifest in the small-amplitude Bogoliubov fluctuations about the current-carrying condensate. Starting from a mean-field condensate, we include Bogoliubov fluctuations~\cite{Fetter-AnnalsOfPhysics} and find the eigenenergies of the quasiparticle excitations. With decreasing repulsion or trap rotational frequency, an energetic instability~\cite{Fetter-RMP, WuNiu-ExcitationStability} can appear in the system via excitations that decrease the angular momentum of the system by one unit; the system can lower its energy by exciting these quasiparticles. Moreover, we find a dynamical instability for sufficiently attractive interactions, where the quasiparticle eigenenergy becomes complex~\cite{Fetter-RMP, WuNiu-ExcitationStability} and the system is driven exponentially rapidly in time away from the initial state. For a system to evolve due to an energetic or a dynamical instability, the presence of dissipation is necessary in order to remove energy and angular momentum; in this paper, we do not include dissipative effects, but will in a future publication. With a knowledge of the instabilities, we then study simple ground states that encompass the underlying physics, consisting of the two lowest-lying single-particle states. (Another example of how an instability indicates the presence of a lower-energy metastable ground state is given in Ref.~\cite{BaharianBaym-VortexFluctuations} where we studied a rapidly rotating trapped Bose gas in the lowest Landau level with a vortex at the center of the trap.)

		For the system to feel the presence of the trap, the trapping potential must break the rotational symmetry. We describe the coupling of the system to the container by an asymmetric ``disorder" potential stationary in the frame rotating at angular velocity $\Omega$. Within mean-field theory, we determine the stationary states of the condensate formed from the single-particle states $\ket{0}$ and $\ket{1}$ and find that for sufficiently large interaction strengths, dependent on the disorder potential, the system exhibits a non-adiabatic response~\cite{Mueller-Hysteresis, WuNiu-LandauZenerTunneling} to variations of the rotation frequency, even if $\Omega$ is changed arbitrarily slowly. This behavior arises from the presence of multiple minima in the energy landscape (separated by a maximum or saddle-point which represents an unstable mode) and is characterized by the appearance of a \emph{swallow-tail loop} in the lowest-lying adiabatic energy level and a fold-over in the occupation probability of the corresponding state as functions of the rotation frequency (see \Fig{energyLoop} below). The swallow-tail loop implies that the response of the system to external rotation exhibits hysteresis~\cite{Mueller-Hysteresis}.

		Moreover, we show that the quasi-one-dimensional Bose-Einstein condensate in a rotating annulus can be mapped onto the problem of a condensate trapped in a double-well potential with Josephson tunneling between the two wells. Therefore, macroscopic quantum phenomenon of self-trapping in double wells~\cite{Smerzi-CoherentDoubleWellTunneling, Milburn-CoherentDoubleWellTunneling} also appears in such rotating Bose gases, where the system acquires a non-zero time-averaged population difference between the two components. The onset of self-trapping, which is a steady-state population imbalance, exactly corresponds to the behavior of the energy levels discussed above.

		We briefly note related theoretical studies in similar toroidally trapped systems: Bose condensates with dipolar interparticle interactions~\cite{BoseInRing} which induce an effective double-well Josephson junction, leading to self-trapping~\cite{DipolarBECs-AbadJezek}; Bose-Einstein condensates with a modulated, spatially-dependent scattering length~\cite{ModulatedScattering}; and hollow pipe optical waveguides with an azimuthally modulated refractive index which generates an effective double-well potential configuration~\cite{Waveguide-Malomed}.

		In Sec.~\ref{Sec-Stability}, we discuss the stability regime of the condensate by studying the energies of the Bogoliubov excitations. We then analyze the energy landscape of the two-mode system in Sec.~\ref{Sec-TwoMode} and demonstrate a swallow-tail loop in the energy of the ground state. We construct a mean-field description of this system in the presence of the disorder potential in Sec.~\ref{Sec-MeanField}. The appearance of swallow-tail loops and cusps in the energy levels and their relation to extrema in the energy landscape are studied in Sec.~\ref{subSec-SwallowTail}. Finally, in Sec.~\ref{subSec-SelfTrapping}, we discuss the connection of this system to a trapped condensate tunneling in a double-well potential, and the corresponding connection of self-trapping in the double-well system to the behavior of the adiabatic energy levels discussed in the previous subsection.

	\section{Stability of the ground state}
		\label{Sec-Stability}
		For a sufficiently thin annulus, the radial and axial excitations are frozen out, and the angle around the ring becomes the only effective degree of freedom. The normalized non-interacting single-particle eigenstates of this system are $\varphi_l(\theta) = \bra{\vect{r}}l\rangle = e^{\mathi l \theta}/\sqrt{2 \pi R}$ with eigenenergies $\epsilon_l = (\hbar l)^2 / 2mR^2$ where $\hbar l$ is the angular momentum and $\vect{r}=(R,\theta)$ is the position vector. The Hamiltonian in the laboratory frame is
		\beq
			\mathcal{H} = \sum_j \frac{(\hbar j)^2}{2mR^2} \, a^\dagger_j a_j + \frac{1}{2} \, \frac{g}{V} \sum_{j,k,m} a^\dagger_{j-m} a^\dagger_{k+m} a_k a_j
			 \label{ringHamiltonian}
		\eeq
		where $a_j$ is the annihilation operator for a particle of angular momentum $\hbar j$, and $g = 4\pi\hbar^2a/m$ is the two-body contact interaction strength with $a$ the $s$-wave scattering length. The Hamiltonian in the frame rotating at $\Omega$ (denoted by a prime) can be written as~\cite{UedaLeggett}
		\begin{align}
			\mathcal{H}^\prime = &- N\frac{\hbar\Omega^2}{2\Omega_0} + \hbar\Omega_0 \sum_j \frac{1}{2} \Big(j-\tfrac{\Omega}{\Omega_0}\Big)^2 \, a^\dagger_j a_j\notag \\
					&+ \frac{1}{2} \, \frac{g}{V} \sum_{j,k,m} a^\dagger_{j-m} a^\dagger_{k+m} a_k a_j
			\label{ringHamiltonianPeriodic}
		\end{align}
		where $\Omega_0 = \hbar/mR^2$ is the characteristic scale of rotation in the system. The non-interacting single-particle energy levels of $\mathcal{H}^\prime$, depicted in \Fig{SPlevels}, are periodic in $\Omega$. At this stage we do not include the disorder potential.
		\bfig[t]
			\centering
			\includegraphics[scale=0.4675]{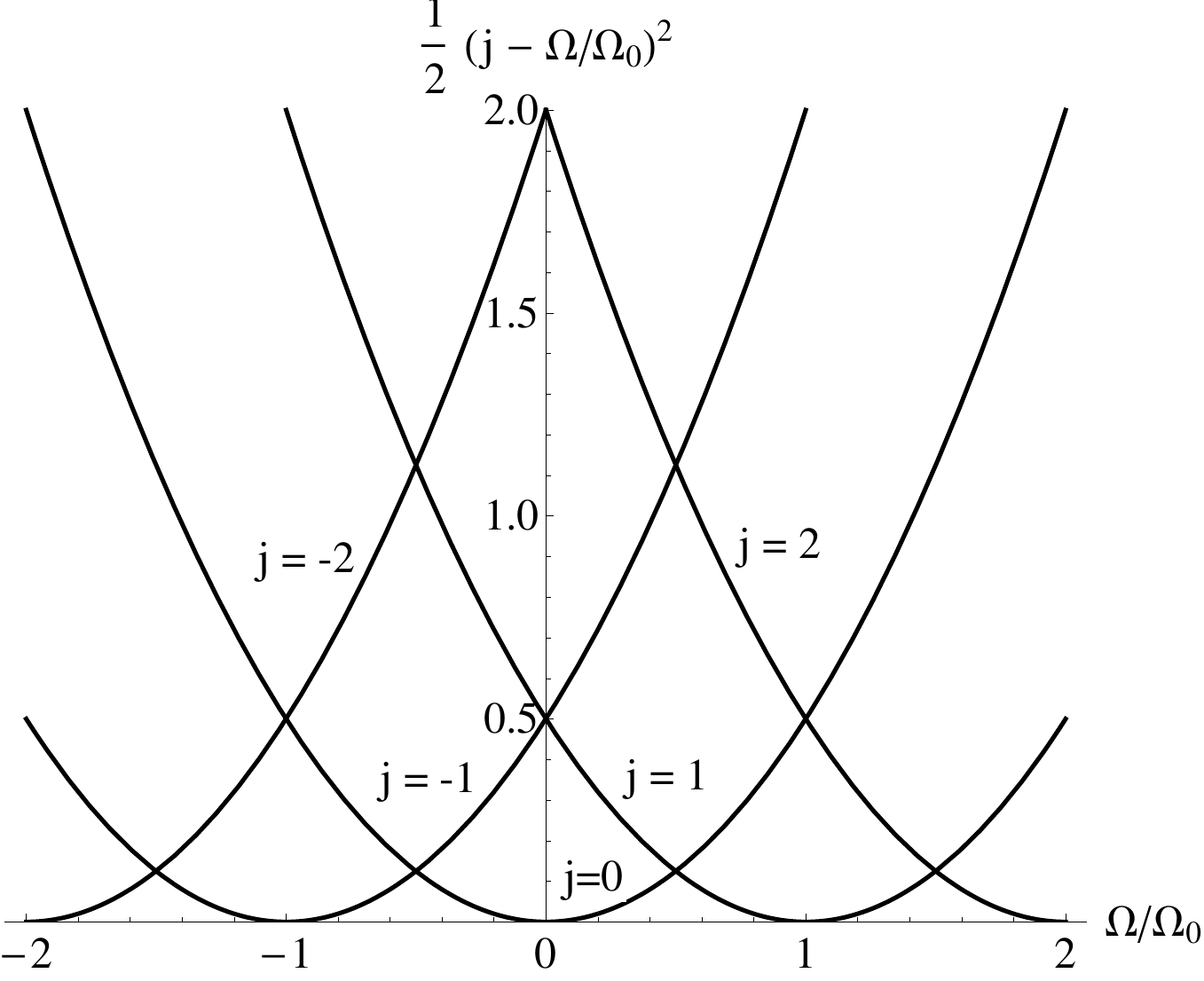}
			\caption{Single-particle energy levels in the rotating frame, measured in units of $\hbar\Omega_0$, as functions of $\Omega$. \label{SPlevels}}
		\efig

		In the laboratory frame, the condensate $\psi_c(\theta, t)$ obeys the time-dependent Gross-Pitaevskii (GP) equation
		\beq
			\mathi\hbar \, \partial_t \psi_c(\theta, t) \! = \! \bigg[ \! - \! \frac{\hbar^2}{2mR^2} \, \frac{\partial^2}{\partial \theta^2} +\frac{g}{\pi r^2_0} \abs{\psi_c(\theta, t)}^2\bigg] \psi_c(\theta, t).
		\eeq
		To determine the stability of the system, we construct the normal modes of the condensate by perturbing the system around the stationary solution $\psi_c(\theta, t) = e^{-\mathi \mu t / \hbar} \, \psi_c(\theta)$, where $\mu$ is the chemical potential. We expand the condensate wave function in terms of time-dependent modes with angular momentum $\nu$ measured relative to the condensate by writing
		\beq
			\psi(\theta, t) = e^{-\mathi \mu t / \hbar} \big[ \psi_c(\theta) + \delta\psi(\theta, t) \big]
		\eeq
		where
		\beq
			\delta\psi(\theta, t) = 	e^{\mathi S(\theta)} \sum_{\nu \neq 0} \big[u_\nu \, \varphi_\nu(\theta) \, e^{-\mathi \epsilon_v t / \hbar} - v^\ast_\nu \, \varphi^\ast_\nu(\theta) \, e^{\mathi \epsilon_v t / \hbar}\big]
			\label{modeExpansion}
		\eeq
		with $\epsilon_v$ the eigenenergies, $S(\theta)$ the phase of $\psi_c(\theta)$, and $u_\nu$ and $v_\nu$ complex numbers to be determined. (Following Fetter's notation~\cite{Fetter-RMP}, we explicitly take the phase of the condensate out of the sum, whereas other authors include the exponential factor in the definition of excitation wave functions~\cite{FiniteTBEC-GPS}.) From now on, for brevity, we measure angular momentum in units of $\hbar$, time in units of $\Omega^{-1}_0$, energy in units of $\hbar\Omega_0$, and define the dimensionless parameters $\eta = mRg / 2 \pi^2 \hbar^2 r^2_0 = 2aR / \pi r^2_0$ and $\bar\Omega = \Omega/\Omega_0$.

		We focus, in particular, on the lowest-energy single-vortex state, with a condensate of $N_c$ atoms in the state $\ket{1}$, for which the GP equation implies that $\mu = \tfrac{1}{2} + \eta N_c$. The modes are described by the two coupled equations~\cite{Fetter-AnnalsOfPhysics}
		\beq
			\begin{pmatrix}
				\nu + \big[\frac{1}{2}\nu^2 + \eta N_c\big] & - \eta N_c \\
				\eta N_c & \nu - \big[\frac{1}{2}\nu^2 + \eta N_c\big]
			\end{pmatrix}
			\begin{pmatrix}
				u_\nu \\
				v_\nu
			\end{pmatrix}
			= \epsilon_\nu
			\begin{pmatrix}
				u_\nu \\
				v_\nu
			\end{pmatrix}
			\label{BogoliubovEquations}
		\eeq
		from which we find the eigenenergy
		\beq
			\epsilon_\nu = \nu + \abs{\nu}\sqrt{\tfrac{1}{4}\nu^2 + \eta N_c} \, .
			\label{ExcitationEnergy}
		\eeq
		Note that for $\tfrac{1}{4}\nu^2 + \eta N_c < 0$, these energies are complex, indicating that the condensate is dynamically unstable~\cite{Fetter-RMP, WuNiu-ExcitationStability}.

		The oscillations of the condensate can also be pictured in second-quantization as quasiparticle excitations of the condensate. In the usual second-quantized Bogoliubov formalism, the coherence factors are given in terms of $\epsilon_\nu$~\cite{Fetter-normalization, Fetter-RMP} by
		\begin{align}
			\abs{u_\nu}^2 &= \frac{1}{2}\left(\frac{\frac{1}{2}\nu^2 + \eta N_c}{\abs{\nu}\sqrt{\tfrac{1}{4}\nu^2 + \eta N_c}} + 1\right) \\
			\abs{v_\nu}^2 &= \frac{1}{2}\left(\frac{\frac{1}{2}\nu^2 + \eta N_c}{\abs{\nu}\sqrt{\tfrac{1}{4}\nu^2 + \eta N_c}} - 1\right),
			\label{uv}
		\end{align}
		and the excitation energy in the rotating frame of a quasiparticle carrying $\nu$ units of angular momentum relative to the condensate becomes
		\beq
			\epsilon^\prime_\nu(\bar\Omega) = \nu (1 - \bar\Omega) + \abs{\nu}\sqrt{\tfrac{1}{4}\nu^2 + \eta N_c} \, .
		\eeq
		Self-consistency dictates that $N_c + \sum_{\nu \neq 0} \abs{v_\nu}^2 = N$.

		\bfig[b]
			\centering
			\includegraphics[scale=0.475]{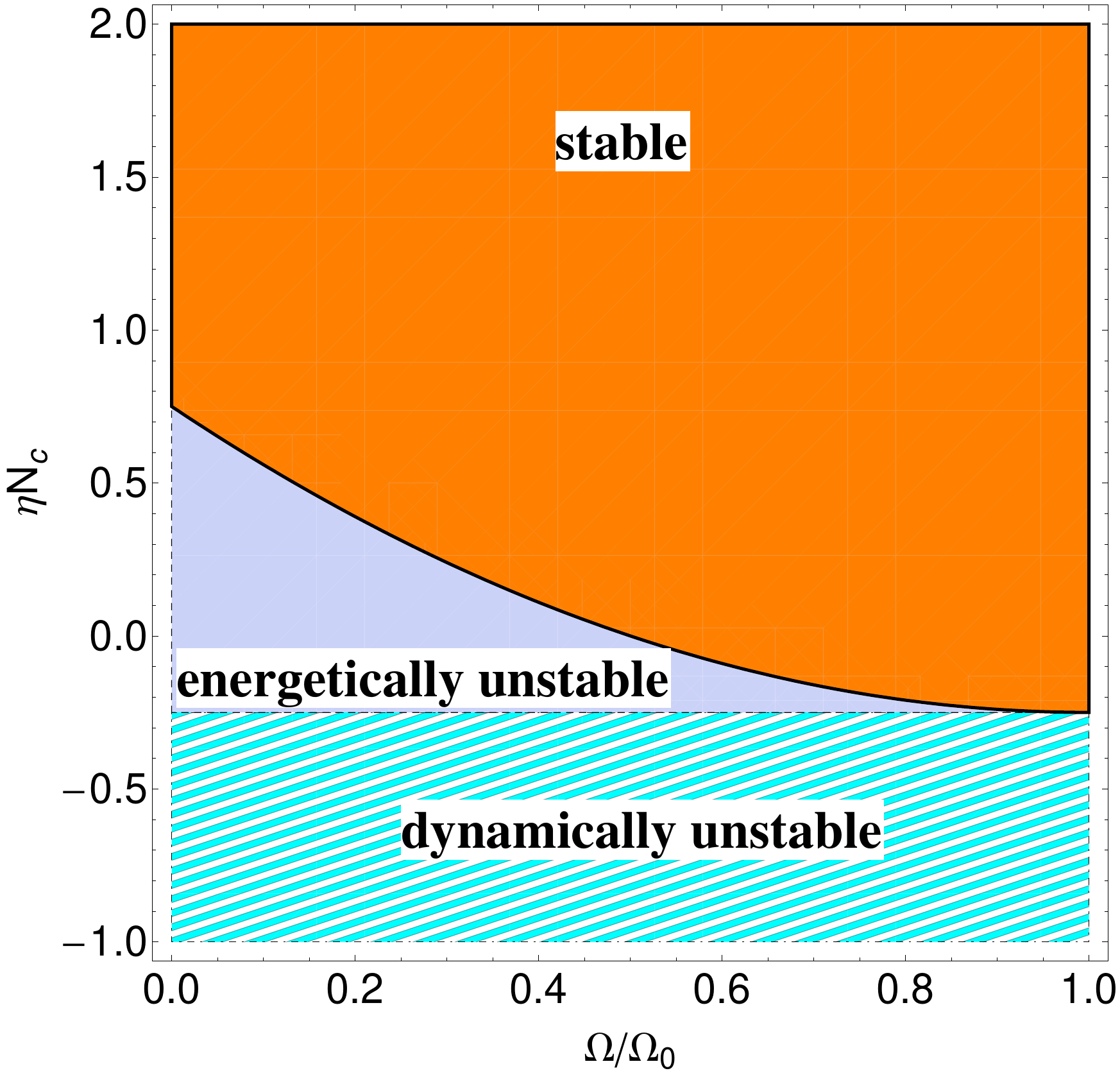}
			\caption{(Color online) Stability phase diagram in the rotation rate -- interaction strength plane, for the $\nu=-1$ normal mode of a condensate with one unit of angular momentum per particle. Energetic instabilities are caused by excitations with negative energy, whereas those with complex energies lead to dynamical instabilities. The current experiments in Refs.~\cite{CurrentDecay-Hadzibabic, BECinToroidalTraps-WeakLink} lie in the stable region. \label{stabilityFig}}
		\efig
		We now analyze the stability of the ground state in terms of the normal modes. For weak interactions, $\abs{\eta N_c} \ll \tfrac{1}{4}$, expansion to first order leads to $\epsilon^\prime_\nu(\bar\Omega) \simeq \tfrac{1}{2}\nu^2 + \nu (1 - \bar\Omega) + \eta N_c$. Thus, at $\bar\Omega=0$ and for repulsive interactions, only $\epsilon_{-1} \simeq -\tfrac{1}{2} + \eta N_c$ is negative; the $\nu = -1$ mode is energetically unstable and anomalous, indicating that the correct ground state has lower angular momentum than the original single-vortex state. For attractive interactions, $\epsilon_{-2} \simeq \eta N_c$ is also negative, and the $\nu = -2$ mode is anomalous as well.

		The general stability phase diagram of the $\nu = -1$ mode is shown in \Fig{stabilityFig} in the interaction strength -- external rotation frequency plane. In the hashed region where $\eta N_c < -\tfrac{1}{4}$, the quasiparticle energy is complex. Note that the regions of dynamical instability and energetic instability are in agreement with the arguments in the appendix of Ref.~\cite{WuNiu-ExcitationStability}. In a dynamically unstable mode, where the eigenenergy is complex, one of the two components in \Eqref{modeExpansion} grows exponentially in time while the other decays exponentially. An unstable mode, living around a maximum or a saddle-point in the energy landscape, hints at the existence of a stable lower-energy state, corresponding to a modified condensate. However, a small-amplitude analysis does not, in general, reveal the nature of the new stable state (see, \eg Refs.~\cite{BaharianBaym-VortexFluctuations, BaymPethick-LandauCriticalVelocity} where such modified condensates are explicitly discussed). The solid black line, the solution of $\epsilon^\prime_\nu = 0$, shows the critical values of interaction strength and rotation frequency needed for stability. For a non-interacting system, the $\nu=-1$ mode becomes stable at $\bar\Omega=\tfrac{1}{2}$. Interestingly, faster rotations shrink the energetically unstable region and stabilize this mode even for weakly attractive interactions. At $\bar\Omega=1$, the energetically unstable region completely vanishes, and the gas becomes stable for $\eta N_c > -\tfrac{1}{4}$. As mentioned before, due to the absence of dissipation in our model, the energy is conserved, and the instabilities (although present) fail to change the state of the system into one with lower energy and lower angular momentum.

		In the experiment in Ref.~\cite{CurrentDecay-Hadzibabic}, where $N \sim 8 \times 10^4$ atoms of $^{87}$Rb with unit circulation were held in a ring trap of radius $R \sim 9 \, \mu\textrm{m}$, we find $\eta N \sim 1.8 \times 10^3$. Also, the experiment in Ref.~\cite{BECinToroidalTraps-WeakLink}, with $N \sim 8 \times 10^4$ atoms of $^{23}$Na in a ring trap of radius $R \sim 20 \, \mu\textrm{m}$, has $\eta N \sim 2.9 \times 10^3$. The initial states in current experiments~\cite{CurrentDecay-Hadzibabic, BECinToroidalTraps-WeakLink} are within the stable regime discussed here, far from encountering any energetical or dynamical instabilities~\footnote{The decay of vorticity seen in these experiments involves a jump from a locally stable state to a lower-energy state, mediated by thermal or quantum tunneling, a problem we take up in a future paper.}. However, by suddenly changing the strength of the interparticle interaction from repulsive to sufficiently attractive via a Feshbach resonance, thereby bringing the system from the stable region into the dynamically unstable region, one would be able to investigate experimentally the evolution of the system in the presence of a dynamical instability.

		The above stability analysis was done for an initial condensate in $\ket{1}$. Due to the periodicity of the single-particle energy levels with respect to the external rotation frequency (see \Fig{SPlevels}), we can extend the same arguments easily to condensates in higher angular momentum states. For a condensate in $\ket{j}$, the chemical potential is $\mu = \tfrac{1}{2} \, j^2 + \eta N_c$, and the quasiparticle energies become $\epsilon^\prime_\nu(\bar\Omega) = \nu (j - \bar\Omega) + \abs{\nu}\big(\tfrac{1}{4}\nu^2 + \eta N_c)^{1/2}$. Thus, the anomalous $\nu=-1$ mode (which connects $\ket{j-1}$ and $\ket{j+1}$ to the condensate) becomes stable at $\bar\Omega=j-\tfrac{1}{2}$ for a non-interacting system; its regions of stability in the presence of interactions, for $j-1 < \bar\Omega < j$, are identical to those shown in \Fig{stabilityFig}.

	\section{Two-mode approximation}
		\label{Sec-TwoMode}
		As discussed above, when the energy of the $\nu=-1$ mode becomes negative, the system, condensed in $\ket{1}$, prefers a ground state with smaller angular momentum than $\hbar$ per particle. As shown in \Fig{SPlevels}, over the entire range $0<\bar\Omega<1$, the lowest-lying single-particle states are $\ket{0}$ and $\ket{1}$. The states $\ket{-1}$ and $\ket{2}$ have, in general, much higher energies and are not mixed in with the ground state by weak interactions; only for $\eta N \gtrsim 1$ is the mixing significant at $\bar\Omega \simeq 0$ and $\bar\Omega \simeq 1$. Hence, for sufficiently weak interactions ($\eta N \lesssim 1$), we can keep only $\ket{0}$ and $\ket{1}$ in the description of the system and work in a two-mode approximation with the truncated Hamiltonian in the frame rotating at $\bar\Omega$,
		\beq
			\mathcal{H}^\prime_2 = (\tfrac{1}{2}-\bar\Omega) \, N_1 + \tfrac{1}{2} \eta (N^2_0 + N^2_1 + 4 N_0 N_1)
			\label{TwoModeHamiltonian}
		\eeq
		where $N_j = a^\dagger_j a_j$. The eigenstates of $\mathcal{H}^\prime_2$ are the Fock states
		\beq
			\ket{N_0, N_1} = \frac{1}{\sqrt{N_0! \, N_1!}} \big(a^\dagger_0\big)^{N_0} \big(a^\dagger_1\big)^{N_1} \ket{\textrm{vac}}
		\eeq
		where $\ket{\textrm{vac}}$ is the vacuum.

		\bfig[t]
			\centering
			\includegraphics[scale=0.235]{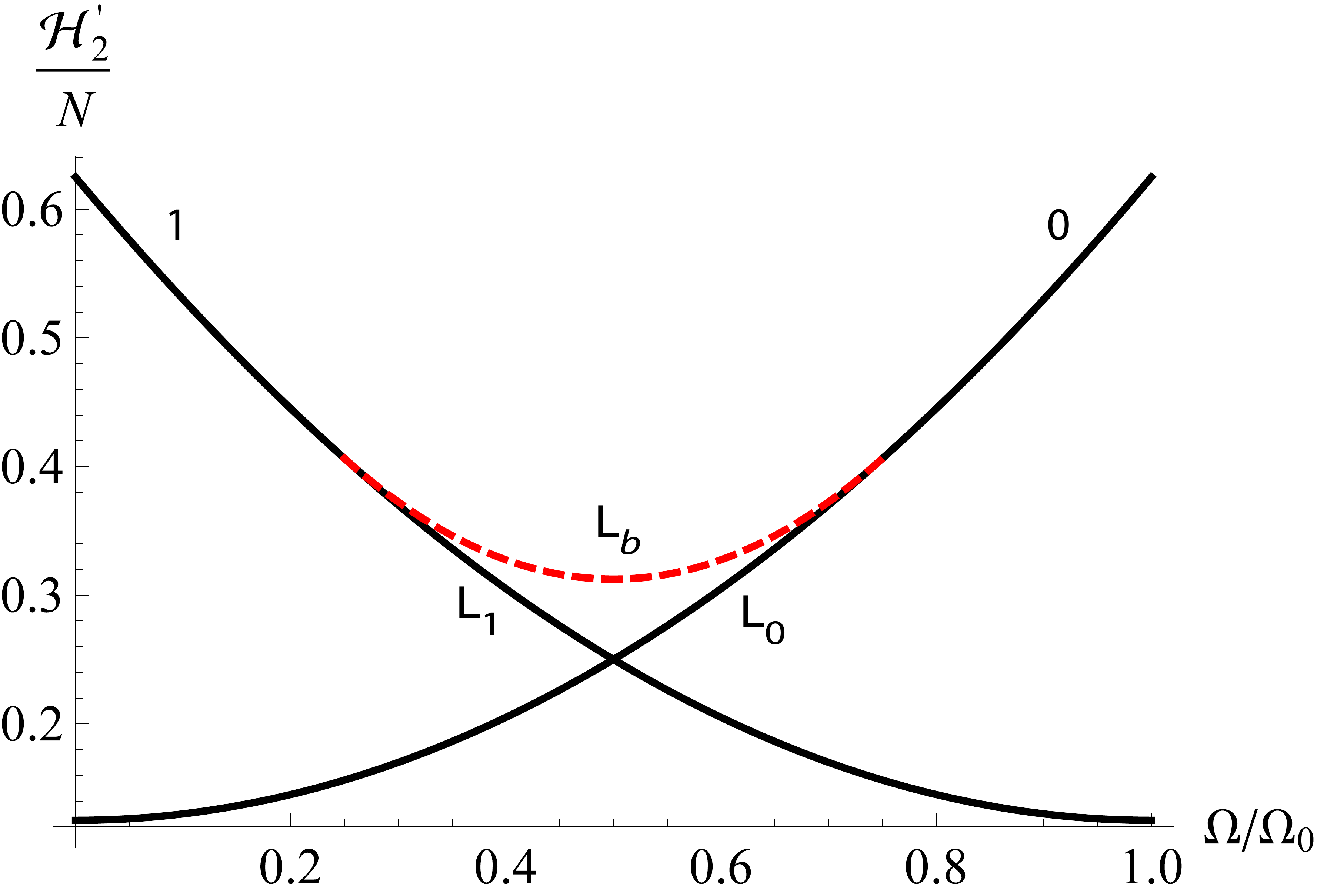}
			\caption{(Color online) The energy per particle, in units of $\hbar\Omega_0$, of the states $\ket{N,0}$ (labeled 0) and $\ket{0,N}$ (labeled 1) and that of the barrier state $\ket{b}$ (dashed line) as functions of the rotation frequency, for $\eta N=1/4$. An energy loop (labeled by $L_b$, $L_0$, and $L_1$) emerges due to the existence of a maximum in the energy landscape.\label{barrierLoop}}
		\efig
		
		This system has been studied extensively in Ref.~\cite{Mueller-Hysteresis}; we briefly recap the results here. Similar to the non-interacting case, in the ground state for $\bar\Omega < \tfrac{1}{2}$, all the particles are condensed into $\ket{0}$, while for $\bar\Omega > \frac{1}{2}$, they are condensed into $\ket{1}$. As one finds by extremizing the energy, the spectrum also acquires an energy maximum when $\abs{\bar\Omega-\tfrac{1}{2}} < \eta N$, corresponding to the state
		\beq
			\ket{b} = \Big| \tfrac{1}{2}N + (\bar\Omega-\tfrac{1}{2}) / 2\eta, \tfrac{1}{2}N - (\bar\Omega-\tfrac{1}{2}) / 2\eta \Big\rangle.
			\label{barrierState}
		\eeq
		This state can be seen as the maximum in \Fig{energyLandscape} for $\bar\Omega=0$ and $\eta N > \frac{1}{2}$; as $\eta N$ decreases below $\frac{1}{2}$, the region in $\bar\Omega$ for which such a maximal state exists shrinks, as depicted by the dashed line in Fig. \ref{barrierLoop} for $\eta N = \frac{1}{4}$. This state acts as a barrier between the two ground states, making it energetically expensive for density modulations (vortex-induced phase slips) to drive the system from one minimum to the other, even as $\bar\Omega$ is varied. Note the loop structure in \Fig{barrierLoop} indicated by the lines labeled $L_b$, $L_0$, and $L_1$, whose effect on the response of the system to changes in $\bar\Omega$ will be discussed in the next section.

	\section{Symmetry breaking in mean-field theory}
		\label{Sec-MeanField}
		We now turn to the question of how the condensate responds dynamically to changes in its rotation rate. In realistic experiments, the potentials felt by the particles are never fully rotationally invariant (see, \eg Ref.~\cite{CurrentDecay-Hadzibabic}); the breaking of rotational invariance changes the way single-particle states of given angular momentum are mixed. In order to couple the system to rotations of the trap, we include in the Hamiltonian a small ``disorder" potential, $\mathpzc{v}(\theta - \Omega t)$, which is stationary in the frame rotating at $\Omega$. Such a potential favors a coherent superposition of states (a single condensate) over a fragmented (Fock) state~\cite{UedaLeggett, Rokhsar-Fragmentation, MHUB-Fragmentation}; we assume the following variational form for the time-dependent two-mode condensate wave function
		\beq
			\ket{\psi_c(t)} = \sqrt{N} \, \big[c_0(t)\ket{0} + c_1(t)\ket{1}\big]
			\label{condensate}
		\eeq
		where $\abs{c_0(t)}^2 + \abs{c_1(t)}^2 = 1$. The time-evolution of the condensate wave function in the rotating frame is governed by the GP equation
 		\beq
			\mathi \partial_t \psi_c(\theta, t) \! = \! \bigg[-\frac{1}{2} \frac{\partial^2}{\partial \theta^2} + \mathi \bar\Omega \frac{\partial}{\partial\theta} + \eta \abs{\psi_c(\theta, t)}^2 + \mathpzc{v}(\theta)\bigg] \psi_c(\theta, t). \notag
		\eeq

		In the two-mode model, we can, with no loss of generality, take $\mathpzc{v}(\theta) = 2\mathpzc{v}\cos\theta$ with $\mathpzc{v}$ real and positive, corresponding to a coupling between the states $\ket{0}$ and $\ket{1}$ of the form $2N\mathpzc{v} \, \textrm{Re}\big[c^\ast_1 c_0\big]$, so that the mean-field Hamiltonian becomes
		\beq
			\frac{\mathcal{H}^\prime_2}{N} = (\tfrac{1}{2}-\bar\Omega) \abs{c_1}^2 + \tfrac{1}{2} \eta N \big(1 + 2\abs{c_0}^2\abs{c_1}^2\big) + 2\mathpzc{v} \, \textrm{Re}\big[c^\ast_1 c_0\big].
			\label{h2}
		\eeq
		With this coupling, the amplitudes obey the two coupled differential equations
		\beq
			\begin{split}
				\mathi \partial_t c_0 &= \eta N \big[2 - \abs{c_0}^2\big] c_0 + \mathpzc{v} c_1, \\
				\mathi \partial_t c_1 &= \eta N \big[2 - \abs{c_1}^2\big] c_1 + \mathpzc{v} c_0  + \big(\tfrac{1}{2} - \bar\Omega\big) c_1.
			\end{split}
			\label{c-Time}
		\eeq
		The angular momentum per particle of the system changes according to
		\beq
			\partial_t\expval{\ell} = -\mathi \partial_t \! \int \!\! d\theta \, \psi^\ast_c(\theta, t) \frac{\partial}{\partial\theta} \psi_c(\theta, t) = 2\mathpzc{v} \, \textrm{Im}\big[c^\ast_1 c_0\big]
		\eeq
		which vanishes, as it should, in the absence of $\mathpzc{v}$ and also when the phase of $c^\ast_1 c_0$ equals $0$ or $\pi$.
		\begin{figure*}[t]
			\centering
			\hspace{-2.5mm}
			\includegraphics[scale=0.16375]{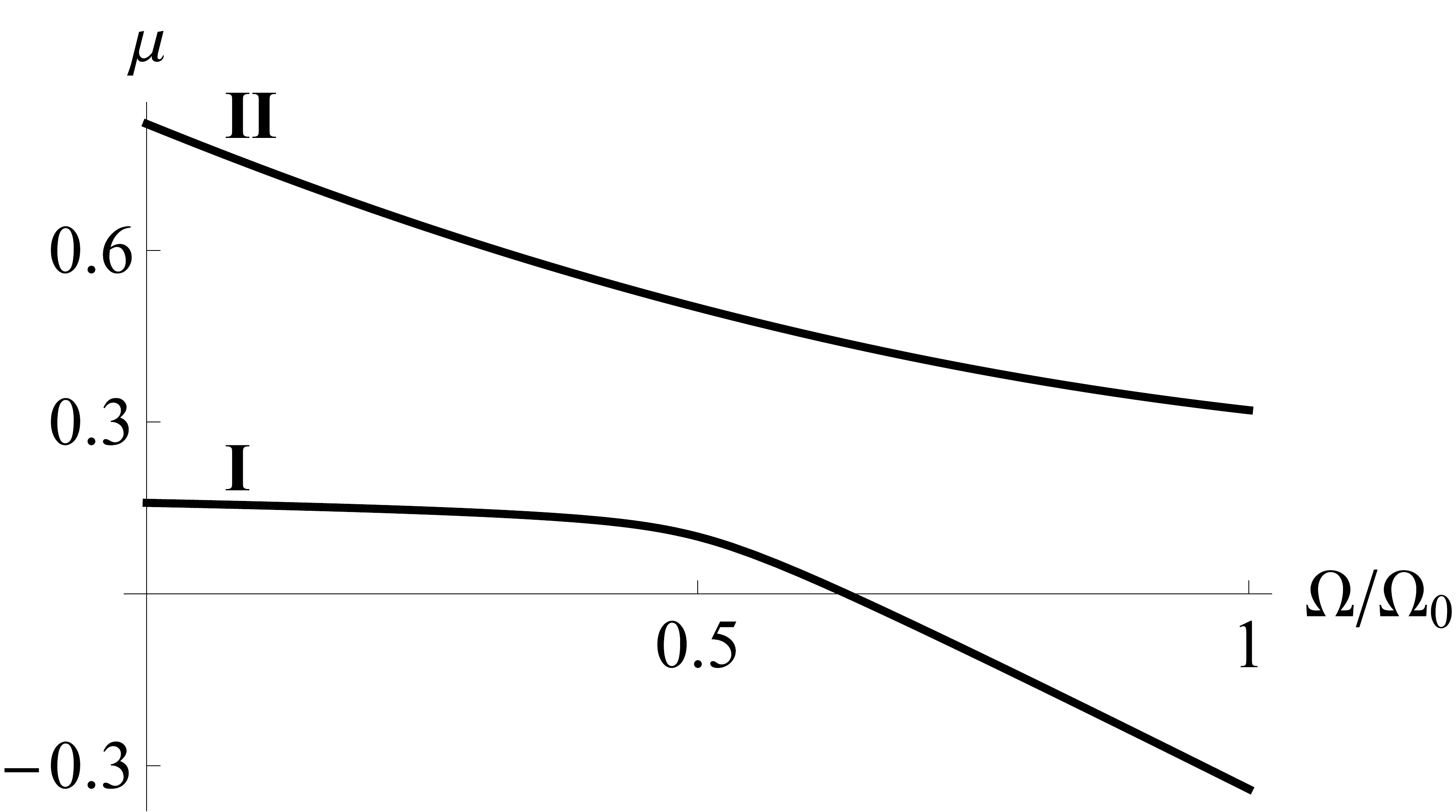}
			\hspace{0.25mm}
			\includegraphics[scale=0.16375]{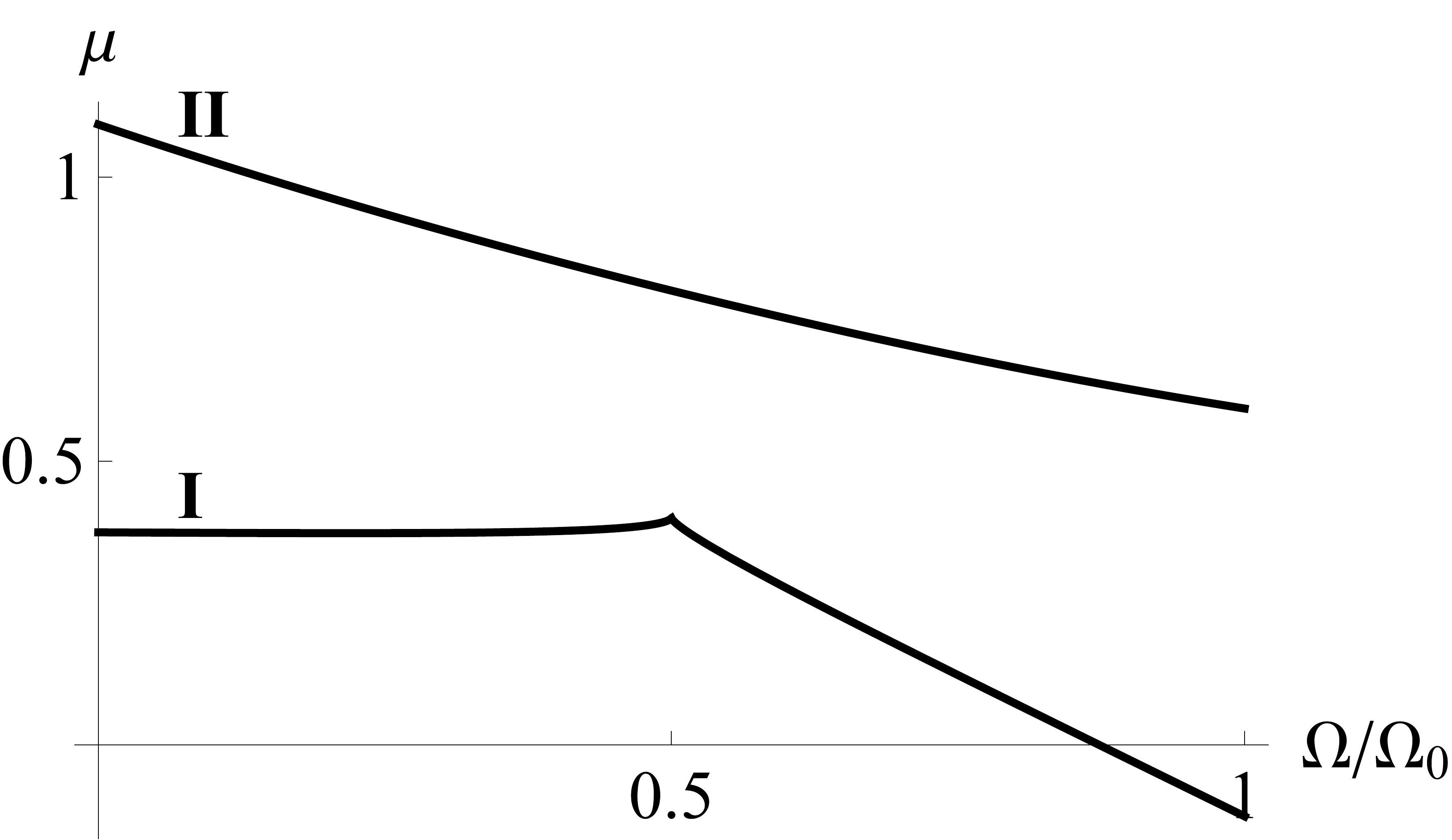}
			\includegraphics[scale=0.16375]{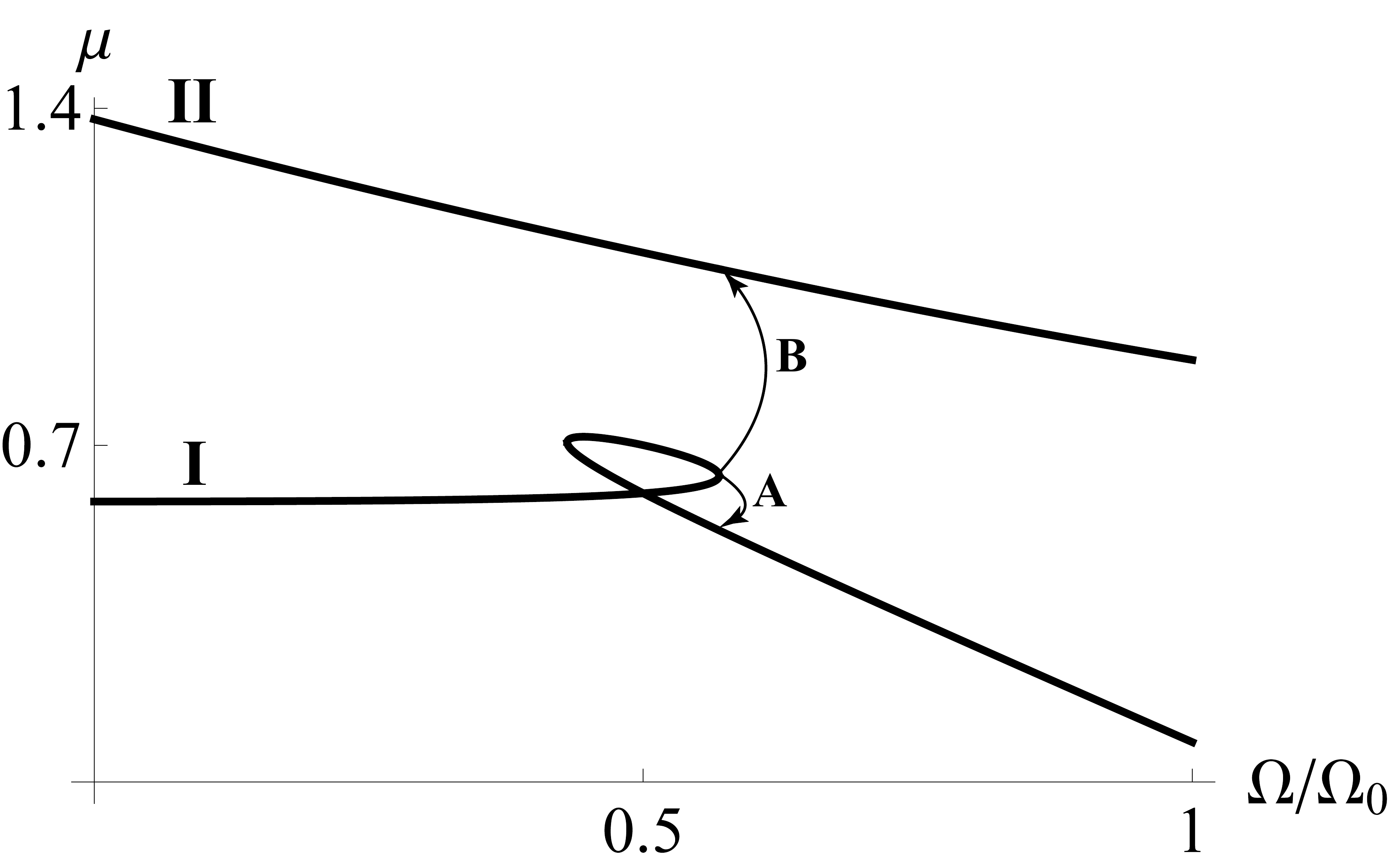} \\
			\vspace{2.5mm}
			\includegraphics[scale=0.1635]{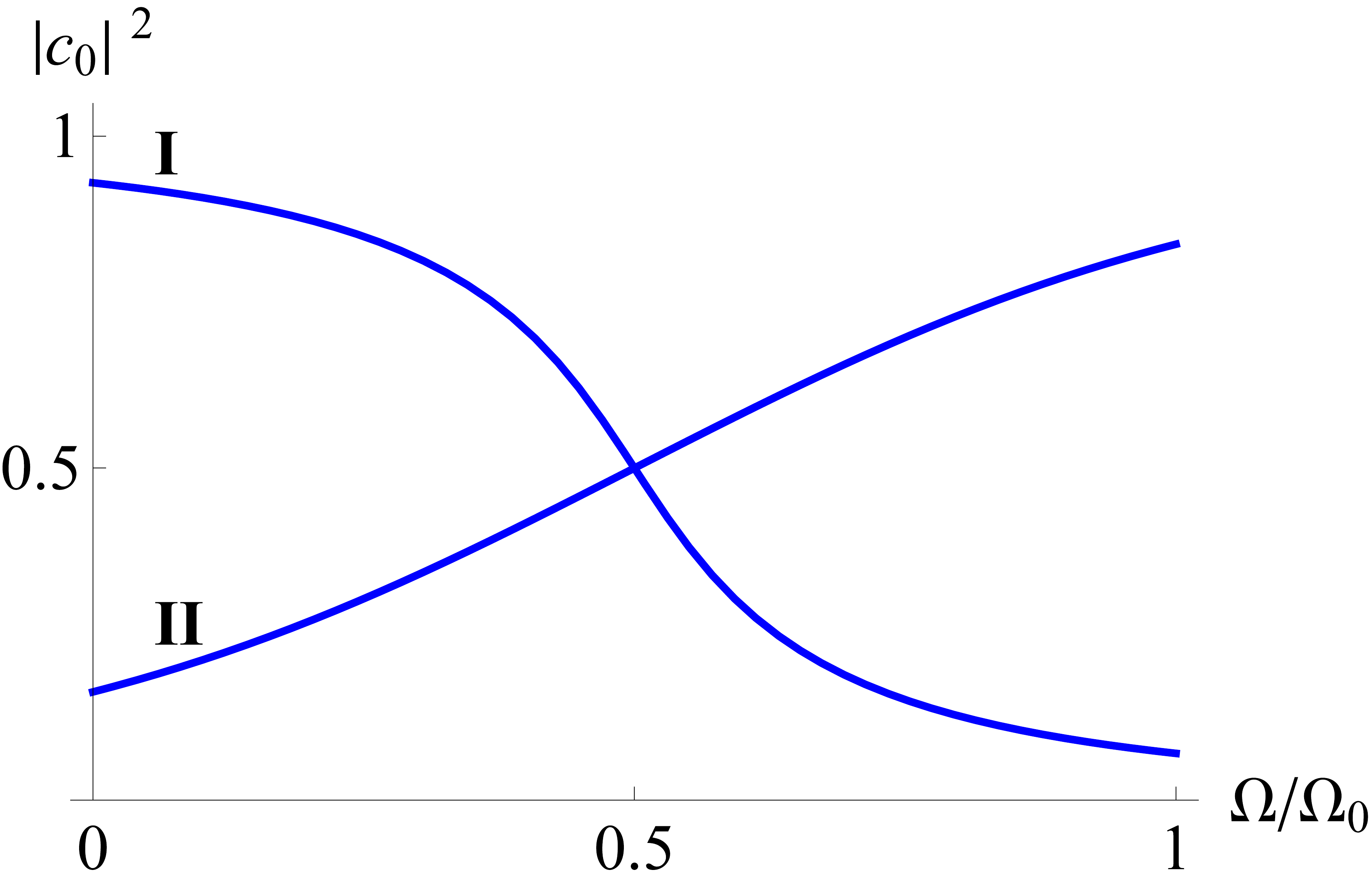}
			\includegraphics[scale=0.1635]{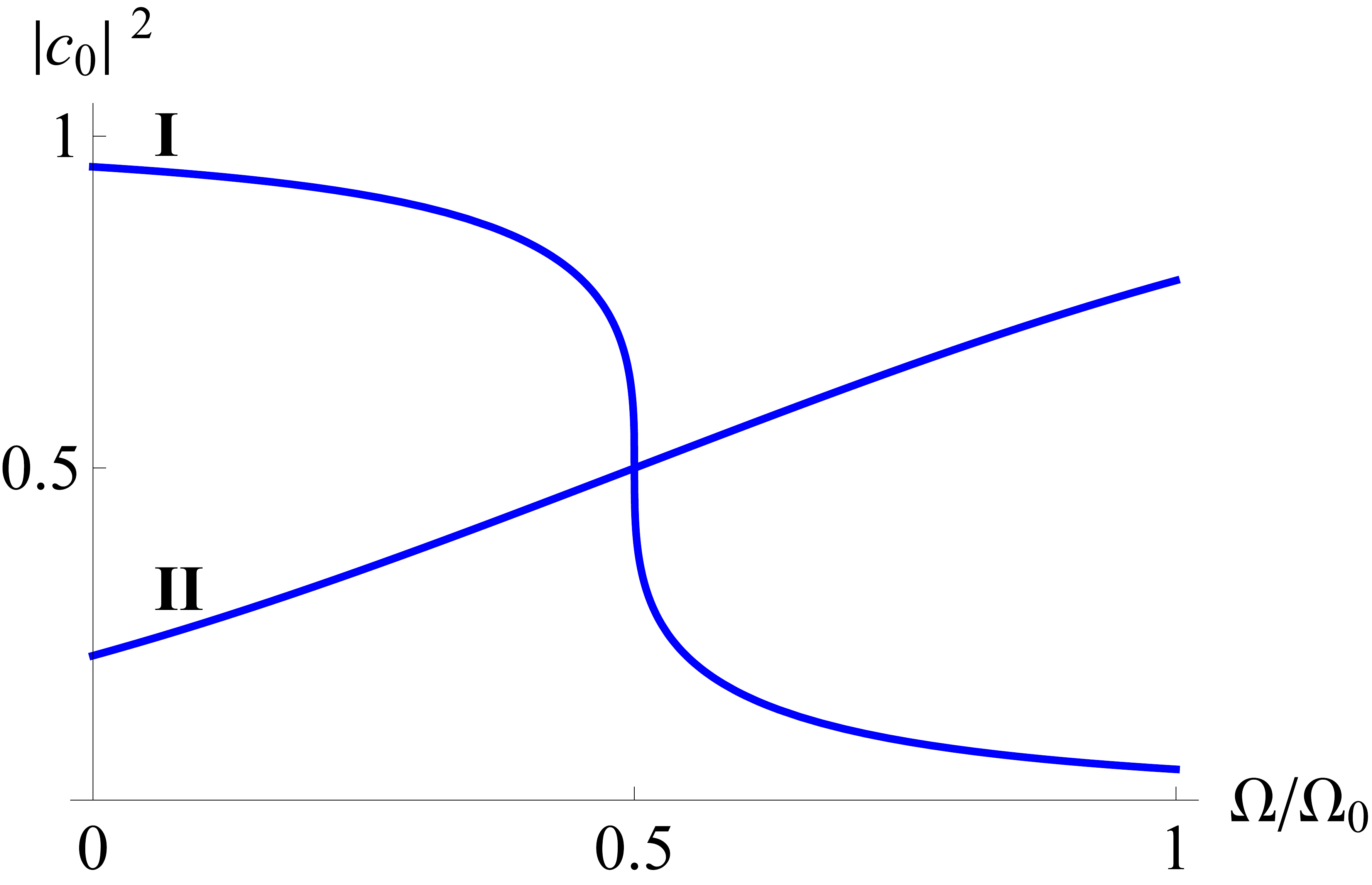}
			\includegraphics[scale=0.1635]{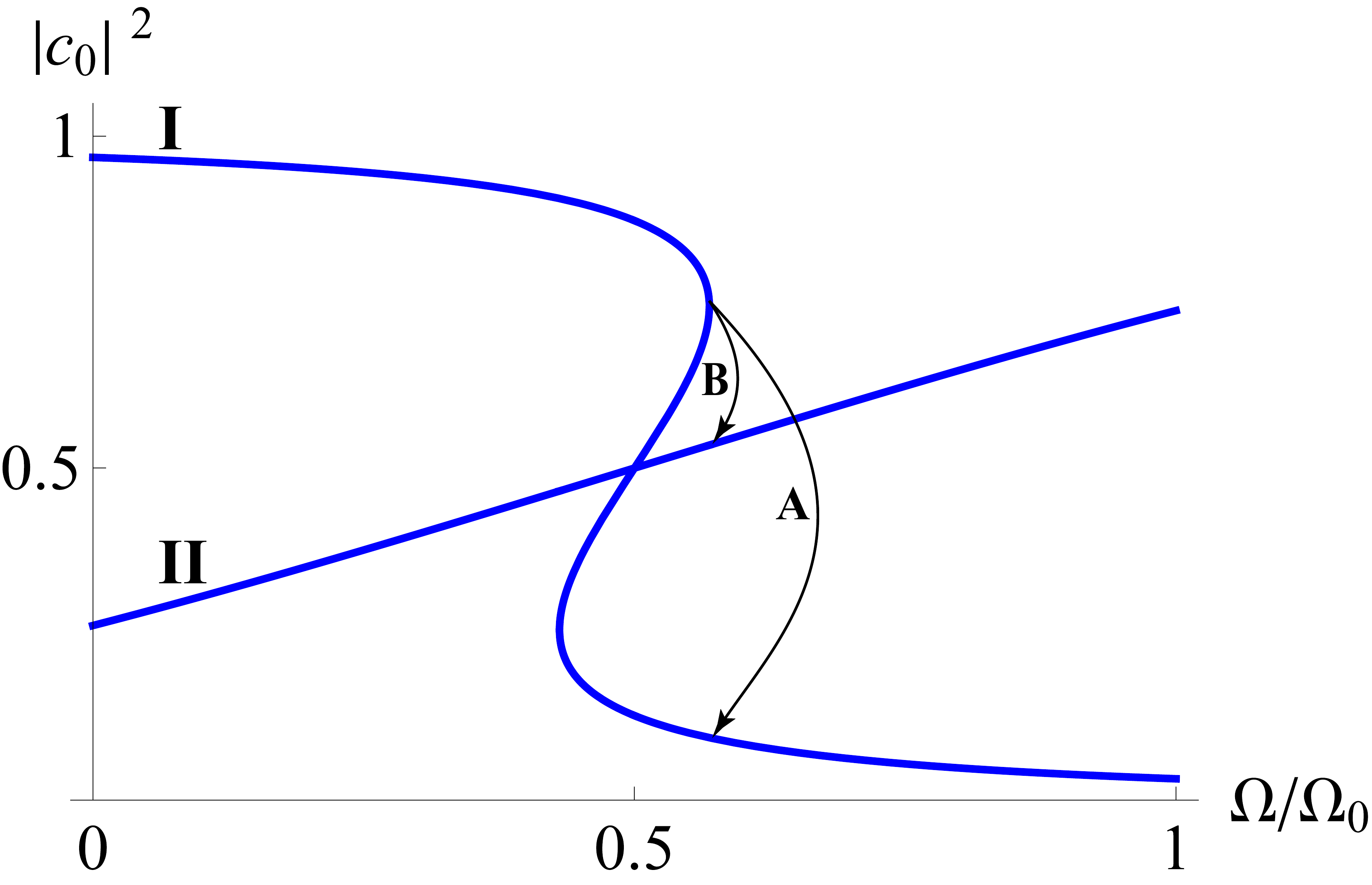}
			\caption{(Color online) Adiabatic energy levels (top panels, in black), measured in units of $\hbar\Omega_0$, and occupation probabilities of $\ket{0}$ (bottom panels, in blue) as functions of $\bar\Omega$ for $\eta N=\mathpzc{v}$ (left column), $\eta N=2\mathpzc{v}$ (middle column), and $\eta N=3\mathpzc{v}$ (right column), with $\mathpzc{v}=1/5$. In all graphs, the lowest energy branch and the corresponding population are indicated by I, and the top energy level and its corresponding population by II. The arrows A and B in the right column indicate the discontinuous change in the population of $\ket{0}$ and the forced tunneling of particles to $\ket{1}$ as $\bar\Omega$ is changed past the folding point.\label{energyLoop}}
		\end{figure*}

		\subsection{Swallow-tail loops}
			\label{subSec-SwallowTail}
			We now ask how the system responds dynamically as the external rotation rate is varied. As we show, the non-linearity inherent in the GP equation leads to forced tunneling between the energy levels of the system in the presence of $\mathpzc{v}$. To see this behavior, we first construct the steady-state solutions of the GP equation in the form $\ket{\psi_c(t)} = e^{-\mathi \mu t}\ket{\psi_c(0)}$ where $\mu$ is the chemical potential, a function of $N$; then, \Eqref{c-Time} gives
			\beq
				\begin{split}
					\mu c_0 &= \Big[\eta N\big(2-\abs{c_0}^2\big)\Big] c_0 + \mathpzc{v} c_1, \\
					\mu c_1 &= \Big[\big(\tfrac{1}{2}-\bar\Omega\big) + \eta N\big(2-\abs{c_1}^2\big)\Big] c_1 + \mathpzc{v} c_0.
				\end{split}
			\eeq
			The occupation probabilities of $\ket{0}$ and $\ket{1}$ as functions of $\mu$ are
			\beq
				\abs{c_0}^2 = \frac{\big(\tfrac{1}{2}-\bar\Omega\big) + \eta N - \mu}{\big(\tfrac{1}{2}-\bar\Omega\big) + 2 (\eta N - \mu)} = 1 - \abs{c_1}^2.
				\label{c0c1mu}
			\eeq
			The eigenstates, then, have the form
			\beq
				\ket{\textrm{I}} =
				\begin{pmatrix}
					\abs{c_0} \\
					-\abs{c_1}
				\end{pmatrix}
				, \quad\quad
				\ket{\textrm{II}} =
				\begin{pmatrix}
					\abs{c_0} \\
					\abs{c_1}
				\end{pmatrix}
			\eeq
			where $\ket{\textrm{I}}$ denotes the ground state branch and $\ket{\textrm{II}}$ the excited state branch, since a phase difference of $\pi$ between $c_0$ and $c_1$ minimizes the coupling energy $2N\mathpzc{v} \, \textrm{Re}\big[c^\ast_1 c_0\big]$ whereas a phase difference of $0$ maximizes it.

			The chemical potential (the ``adiabatic energy level" in the sense of Refs.~\cite{WuNiu-LandauZenerTunneling, WuNiu-ExcitationStability}) is found from the determinant
			\beq
				\begin{vmatrix}
					\eta N\big(2-\abs{c_0}^2 \! \big) - \mu & \mathpzc{v} \\
					\mathpzc{v} & \big(\tfrac{1}{2}-\bar\Omega\big) + \eta N\big(2-\abs{c_1}^2 \! \big) -\mu
				\end{vmatrix}
				=0
			\eeq
			together with Eqs.~\eqref{c0c1mu}. The result is a fourth-order equation for $\mu$, with two to four real solutions depending on the values of $\eta N / 2\mathpzc{v} \equiv \Lambda$ and $\bar\Omega$. The chemical potential is simply related to the energy per particle in the rotating frame by
			\beq
				E^\prime = \mu - \tfrac{1}{2} \eta N \Big(1 + 2 \abs{c_0}^2 \abs{c_1}^2\Big).
				\label{E-mu}
			\eeq

			The energy levels corresponding to states $\ket{\textrm{I}}$ and $\ket{\textrm{II}}$, in general, exhibit an avoided crossing as a function of $\Omega$ due to the presence of disorder. Since $\mu$ and $E^\prime$ are simply related by \Eqref{E-mu}, the physical content of their corresponding plots is identical. To illustrate the physics, we plot the behavior in terms of $\mu$ since it is graphically clearer. The upper panels of \Fig{energyLoop} show the real solutions for $\mu$ as functions of $\bar\Omega$ for selected values of $\eta N$ and $\mathpzc{v}$. As shown in the figure, and as we will prove at the end, for $\Lambda<1$, the two energy levels have an avoided crossing at the critical value of the rotation rate $\bar\Omega_c = \frac{1}{2}$; at $\Lambda=1$, a cusp appears in the lower branch at this frequency; and for $\Lambda>1$, the cusp gives birth to a loop in the lower branch. The loop discussed earlier in Sec.~\ref{Sec-TwoMode} in the absence of the disorder potential (see \Fig{barrierLoop}) evolves into the present loop as the disorder is turned on. Note that at a given rotation frequency, \Fig{barrierLoop} shows either two or three states, while \Fig{energyLoop} shows two or four states; the extra state arises from mixing of the upper maximum-energy state with lower-energy states (not shown in \Fig{barrierLoop}).

			As seen in the lower panels of \Fig{energyLoop}, the derivative of the occupation, $\abs{c_0}^2$, of $\ket{0}$ with respect to $\bar\Omega$ diverges as the cusp appears in the lowest energy band, and the occupation folds over itself (the characteristic S shape seen in fold catastrophes~\cite{Berry-Catastrophe, ODell-LoopsInCavity, ODell-BoseImpurityDoubleWell}) as the loop emerges for $\Lambda>1$. The swallow-tail loop indicates hysteresis~\cite{Mueller-Hysteresis} and a lack of adiabatic evolution with $\bar\Omega$~\cite{WuNiu-LandauZenerTunneling, LiuWuNiu-AdiabaticEvolution, Loop-PethickSmith, WuNiu-ExcitationStability, BlochAltman-Loop} in a condensate in an annulus.

			To see the physics of the swallow-tail loop, imagine that we prepare the system, with $\Lambda>1$, on the lower branch at $\bar\Omega=0$ and very slowly increase $\bar\Omega$ to avoid any tunneling to the other branch. The system will, then, follow this branch adiabatically until the point where the branch terminates and folds back on itself (at $\bar\Omega>\tfrac{1}{2}$). Upon further increase of $\bar\Omega$, the system is forced to make a discontinuous jump either to the lower part of branch I (indicated by the arrow A in \Fig{energyLoop}) or to the upper branch II (indicated by the arrow B). Similarly, the occupation probability of $\ket{0}$ adiabatically follows the change in $\bar\Omega$ until the branch starts to fold over itself, at which point a sudden change in the population of that state becomes inevitable with further increase of the rotation frequency, as indicated by the arrows in \Fig{energyLoop}. In other words, a fraction of the particles in $\ket{0}$ are forced to tunnel to $\ket{1}$. Sweeping $\bar\Omega$ in the other direction forces a similar behavior on the system as well.

			The folded-over section of branch I of the occupation probability and the respective top part of the swallow-tail loop of the lowest-energy level (in the right column of \Fig{energyLoop}) are inaccessible through a sweep of $\Omega$ and correspond to unstable states. In direct analogy with the barrier state discussed earlier, they indicate the presence of more than one minimum in the energy landscape, separated by a maximum or a saddle-point~\cite{Mueller-Hysteresis, ODell-LoopsInCavity}. As we show in the next subsection, the appearance of the cusp (along with the corresponding divergence of $\partial \! \abs{c_0}^2 \!\! / \partial \bar\Omega$ at $\bar\Omega_c$) and the swallow-tail loop (along with the corresponding fold-over in the level population) are related to the macroscopic quantum phenomenon of self-trapping or self-locked population imbalance.

		\subsection{Self-trapping}
			\label{subSec-SelfTrapping}
			\begin{figure*}[t]
				\centering
				\includegraphics[scale=0.1175]{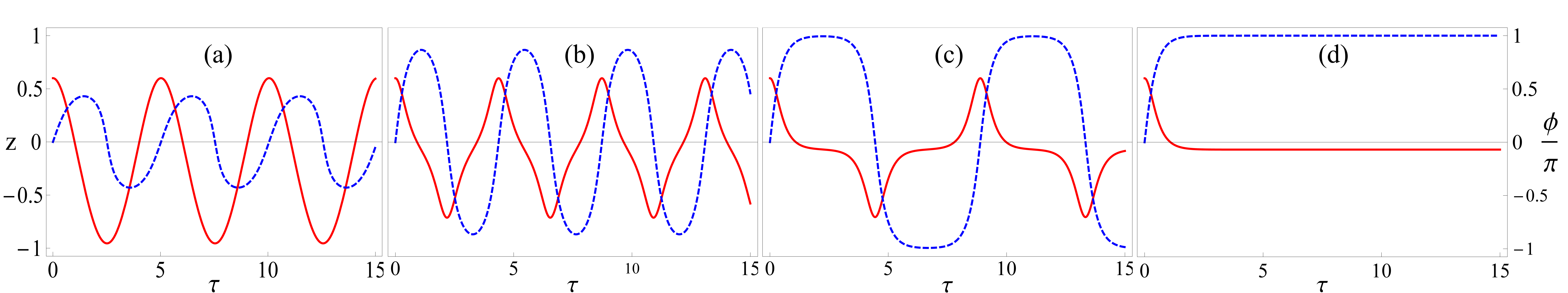}
				\caption{(Color online) Population difference $z$ (solid line, in red) and phase difference $\phi/\pi$ (dashed line, in blue) as functions of $\tau$ for $\Delta E = 1/2$ and (a) $\Lambda = 1$ below the critical value, (b) $7.75$, (c) $8.2365$ just below the critical value, and (d) $8.2374$ the critical value. The initial conditions are $z(\tau=0) = 0.6$ and $\phi(\tau=0) = 0$. Note how the oscillatory behavior of $z(\tau)$ changes from purely harmonic to anharmonic as $\Lambda$ increases; finally, $z(\tau)$ and $\phi(\tau)$ both become time-independent as $\Lambda$ reaches the critical value. \label{populationImbalance}}
			\end{figure*}
			Interesting quantum phenomena, including Josephson effects analogous to those in superconductors and also chaotic dynamical behaviors, arise from the dynamical behavior of the macroscopic phase difference between the two components of the time-dependent condensate~\eqref{condensate}. To see this connection, we recast the time-evolution equations~\eqref{c-Time} in terms of the population difference $z = \abs{c_0}^2-\abs{c_1}^2$ (where $-1 \le z \le 1$) and the phase difference $\phi = \alpha_0 - \alpha_1$ between the two constituent states, where $\alpha_j = \arg[c_j]$. Then, we find
			\begin{align}
				\partial_\tau z &= -\sqrt{1-z^2} \, \sin\phi \label{ztau} \\
				\partial_\tau \phi &= \Delta E + \Lambda z + \frac{z}{\sqrt{1-z^2}} \, \cos\phi
				\label{phitau}
			\end{align}
			where we rescale the time to $\tau = 2 \mathpzc{v} t$ and set
			\beq
				\Delta E = \big(\tfrac{1}{2}-\bar\Omega\big) / 2\mathpzc{v}.
			\eeq
			These equations are identical to Eqs.~(3a) and (3b) of Ref.~\cite{Smerzi-CoherentDoubleWellTunneling} which describes coherent tunneling between two Bose-Einstein condensates in a double-well potential; therefore, all the results of that paper directly apply to the present system. The two levels, $\ket{0}$ and $\ket{1}$, correspond to the two wells. In the double-well system, an applied DC voltage between the two wells induces tunneling and, therefore, an AC particle current
			\beq
				\mathcal{I} = 2N\mathpzc{v} \, \partial_\tau z
			\eeq
			between the two condensates~\cite{Milburn-CoherentDoubleWellTunneling}. Analogously, in a toroidal trap, an external rotation induces a population transfer between the two levels.

			The Hamiltonian~\eqref{h2} written in terms of $\phi$ and $z$ is given, to within a constant term, by
			\beq
				\frac{\mathcal{H}_2^\prime}{N\mathpzc{v}} = -\Delta E \, z - \tfrac{1}{2}\Lambda z^2 + \sqrt{1-z^2} \, \cos\phi
			\eeq
			and is a conserved quantity. Given that the dynamics is Hamiltonian, the quantum analog of the Poincar\'e recurrence theorem holds~\cite{QuantumRecurrenceTheorem}, and, therefore, the system is inherently periodic in time.

			The time evolution of the system, calculated numerically, is shown in \Fig{populationImbalance} for different $\Lambda$. With increase of $\Lambda$ for a given initial population imbalance, $z(0)$, the oscillations in $z(\tau)$ change from purely harmonic to anharmonic and a plateau appears in $z(\tau)$ [the nearly flat part of the curve in \Fig{populationImbalance}(c)]. At a critical value of $\Lambda$, dependent on $z(0)$, the oscillation period becomes infinite, and the population imbalance becomes time-independent at $z(\tau)=z_s$ [see \Fig{populationImbalance}(d)]. For $\Lambda$ larger than this critical value, $z(\tau)$ oscillates in time entirely above $z_s$ (not shown in \Fig{populationImbalance}). Similarly, for fixed $\Delta E$ and $\Lambda$, there exists a critical value $z_c$ of the initial population difference for which the oscillations cease and $z(\tau)$ becomes constant after a finite time (the plateau continues indefinitely). This evolution to a state with a non-zero time-averaged value of $z(\tau)$ (independent of $\Delta E$ and for $\Lambda$ greater than or equal to the critical value discussed above) is the analog of the phenomenon of self-trapping in the double-well system~\cite{Smerzi-CoherentDoubleWellTunneling,Milburn-CoherentDoubleWellTunneling}.

			The condition to develop self-trapping is that the two time derivatives, $\partial_\tau z$ and $\partial_\tau \phi$, vanish simultaneously. From \Eqref{ztau}, this requires $\phi=0$ or $\pi$ (although it appears that the singular point $z=1$ also makes $z(\tau)$ time-independent, the correct solution is actually time-dependent~\cite{Milburn-CoherentDoubleWellTunneling}, as can be deduced from \Eqref{c-Time}, and is thus unacceptable). From \Eqref{phitau}, the steady-state value $z_s$ is given in terms of $\Delta E$ and $\Lambda$ by
			\beq
				\Delta E + \Lambda z_s \pm z_s / \sqrt{1-z^2_s} = 0.
				\label{lockedImbalance}
			\eeq
			Once the system reaches this plateau, it must stay there forever, since the equations of motion are first-order in time. The critical initial population difference $z_c$ that leads to self-trapping can be found from energy conservation. For an initial phase difference, $\phi(0)$, we find
			\begin{align}
				&\Delta E \, z_c + \tfrac{1}{2}\Lambda z^2_c - \sqrt{1-z^2_c} \, \cos\phi(0) \notag \\
				&= \Delta E \, z_s + \tfrac{1}{2}\Lambda z^2_s \mp \sqrt{1-z^2_s} \, .
			\end{align}
			As illustrated in \Fig{populationImbalance}, the stationary self-trapped population imbalance $z_s$ is, in general, non-zero and only vanishes for $\Delta E=0$.

			The steady-state self-trapped solution is, in fact, related to the adiabatic energy levels discussed above. The stationary value $\phi = \pi$ leads to the ground state (branch I in \Fig{energyLoop}), whereas $\phi = 0$ gives the excited state (branch II in \Fig{energyLoop}). We focus first on the point $\bar\Omega=\bar\Omega_c$ at which the cusp and the tip of the swallow-tail loop appear and for which $\Delta E=0$; choosing the minus sign in \Eqref{lockedImbalance} (corresponding to the ground state), we find three solutions
			\beq
				z_s = 0, \pm\sqrt{1-\Lambda^{-2}} \, .
			\eeq
			For $\Lambda<1$, two solutions are complex and unphysical; however, when $\Lambda=1$, all three solutions become degenerate at $z_s=0$; and for $\Lambda>1$, we have three distinct solutions. This change in the number of solutions at the critical point $\Lambda=1$ is another indication of the occurence of the catastrophe~\cite{ODell-LoopsInCavity, Berry-Catastrophe} when a swallow-tail loop appears. Moreover, \Eqref{lockedImbalance} also yields
			\beq
				\frac{\partial z_s}{\partial \bar\Omega} = -\frac{1}{\Lambda \big(1-\Lambda^2\big)} = 2 \, \frac{\partial \abs{c_0}^2}{\partial \bar\Omega}
			\eeq
			for $\Delta E=0$. For $\Lambda=1$, this quantity diverges; for $\Lambda<1$, it is negative; and for $\Lambda>1$, it is positive. Together, these two quantities exhibit the exact behavior, with varying $\Lambda$, seen in the bottom panels of \Fig{energyLoop} at $\bar\Omega=\bar\Omega_c$. In general, for non-zero $\Delta E$, finding $z_s$ as a function of $\Omega$ by solving \Eqref{lockedImbalance}, we indeed see the behavior for $\abs{c_0}^2$ depicted in \Fig{energyLoop}. Hence, the ceasing of the coherent oscillation between the two components of the condensate (as the system becomes self-trapped) and the appearance of a cusp or a swallow-tail loop in the lowest-lying adiabatic energy level are in one-to-one correspondence. This proves our previous statement that the critical disorder strength for which a cusp or a loop first appears is $\Lambda = 1$ or, in other words, $\mathpzc{v} = \eta N/2$.

	\section{Summary}
		In this paper, we analyze the stability of a Bose-Einstein condensate in a rotating toroidal trap in terms of the normal modes of small-amplitude deviations from a metastable current-carrying state. We identify regions of energetic and dynamical instabilities in the phase space as functions of the interparticle interaction strength and rotation rate of the trap. Describing the coupling of the system to the rotation of the trap by a symmetry-breaking disorder perturbation to the original Hamiltonian, we investigate the steady-state and the general dynamics of this system in a two-mode mean-field model. We find that in the presence of these disorders, swallow-tail loops appear in the lowest-lying energy band for sufficiently strong interaction strengths and, as the rotation rate is varied, force a sudden, non-adiabatic change in the state of the system and the population of the two constituent states. Finally, we investigate the connection of this system with a system of two condensates tunneling in a double-well potential and find these two systems to have identical dynamics; therefore, the analog of the phenomenon of self-trapping also appears in the system studied in this paper. We calculate the onset and the properties of self-trapping, and show how the steady-state self-trapped states are described in terms of the energy eigenstates.

		The next step, which we discuss in a future publication, is to include the effects of non-zero temperature on the detailed time-dependence of a Bose gas rotating in an annulus. At finite temperature, the angular momentum and energy of the condensate need not be conserved. As a result, thermal, as well as quantum, fluctuations can induce transitions between metastable states, as has been observed experimentally~\cite{CurrentDecay-Hadzibabic}. Furthermore, the coupling of an unstable condensate to the environment would allow it to evolve to a stable configuration.

	\begin{acknowledgements}
		This work was supported in part by National Science Foundation Grants No.~PHY07-01611 and PHY09-69790. Author S.B. would like to thank Alexander Fetter for an illuminating discussion and Brian DeMarco for critical comments on the manuscript and further useful suggestions.
	\end{acknowledgements}

\end{document}